\documentclass{ws-procs9x6}

\newcommand{\be}{\begin{equation}}
\newcommand{\ee}{\end{equation}}
\newcommand{\ba}{\begin{array}}
\newcommand{\ea}{\end{array}}
\newcommand{\bea}{\begin{eqnarray}}
\newcommand{\eea}{\end{eqnarray}}

\begin{document}

\title{EMBEDDED REPRESENTATIONS \\
AND QUASI-DYNAMICAL SYMMETRY%
\footnote
{\uppercase{I}nvited talk: \uppercase{I}nt.\ \uppercase{C}onf.\ on
\uppercase{C}omputational and \uppercase{G}roup-\uppercase{T}heoretical
\uppercase{M}ethods in \uppercase{N}uclear \uppercase{P}hysics
(\uppercase{F}eb.\ 18-21, 2003, \uppercase{P}laya del \uppercase{C}armen,
\uppercase{M}exico, in \uppercase{H}onor of \uppercase{J}erry
\uppercase{P}.\ \uppercase{D}raayer's 60th \uppercase{B}irthday), eds.\
\uppercase{J.\ E}scher, \uppercase{O.\ C}asta\~nos, \uppercase{J.G.\
H}irsch, \uppercase{S.\ P}ittel, \uppercase{G.\ S}toitcheva
(\uppercase{W}orld \uppercase{S}cientific, \uppercase{S}ingapore, 2004).}}

\author{D.~J. ROWE}

\address{Deparment of Physics,
University of Toronto, \\ 
Toronto, Ontario, M5S 1A7, Canada\\ 
E-mail: rowe@physics.utoronto.ca}


\begin{abstract}
This presentation  explains why models
with a dynamical symmetry often work extraordinarily well even in the
 presence of large symmetry breaking interactions.
A model may be a caricature of a
more realistic system with a ``quasi-dynamical" symmetry.
The existence of quasi-dynamical symmetry in physical systems and
its significance for understanding collective dynamics in
complex nuclei is explained in   terms of the precise mathematical concept
of an ``embedded representation".  
Examples are given which exhibit
quasi-dynamical symmetry to a remarkably high degree. 
Understanding this unusual symmetry and why it occurs, is
important for recognizing why dynamical symmetries appear to be much more
prevalent than they would otherwise have any right to be and for
interpreting the implications of a model's successes. 
We indicate when quasi-dynamical symmetry is expected to apply and
present a challenge as to how best to make use of this potentially
powerful algebraic structure.
\end{abstract}

\section{Introduction} 

I intended to talk about vector coherent state theory.  However,
several examples shown by others of what Jerry Draayer
appropriately referred to as an {\it adiabatic coherent mixing\/} of
representations, prompted me to change my topic to a description of
the mathematical structure and physical significance of this potentially
powerful and physically useful concept.

When a simple model is successful at describing a physical system,
there is a temptation to infer that the model has a corresponding
degree of  reality.
However, it is easy to be misled.
This concern led us to investigate why systems
frequently  appear to hold onto a dynamical symmetry in spite of strong
symmetry-breaking interactions.
This  is particular evident in systems
which exhibit a Landau second order transition from a phase with
one apparent symmetry to a phase with a different symmetry.
The outcome was the discovery of quasi-dynamical
symmetry\cite{Carv,RRR,RR88}.

\section{What is quasi-dynamical symmetry?}

It is well known that states of different, but equivalent irreps, of a Lie
algebra (or Lie group) can mix coherently to form new irreps.
For example, if $\{ |\alpha LM\rangle\}$ are states of angular momentum
$L$ and $z$-component $M$, with $\alpha$ distinguishing different states
of the same angular momentum, then the states
\be \{|\Psi_{\kappa LM}\rangle = \sum_\alpha C_{\kappa \alpha} |\alpha
LM\rangle, \; M=-L,\dots , +L\} 
\ee
span another (equivalent) so(3) irrep of angular momentum $L$.
What is remarkable is that, for some Lie algebras, there are linear
combinations of states from similar, but inequivalent, irreps that
actually form a basis for an irrep of the Lie algebra. Such an irrep is
called an {\it embedded representation\/}.  They may seem like
bizarre mathematical oddities but, in fact, embedded representations are
common in physics and underlie the  {\it adiabatic
separation of variables}. We say that a model has a quasi-dynamical
symmetry if its states span a so-called {\it  embedded representation\/}
of a Lie algebra\cite{RRR}.

\medskip\noindent{\bf Definition:}  If $\mathbb{H}$ is the Hilbert space
for a (generally reducible) representation $U$ of a Lie algebra ${\frak
g}$ and $\mathbb{H}_0\subset \mathbb{H}$ is a subspace then, if the
matrix elements of $\frak g$ between states lying in $\mathbb{H}_0$ are
equal to those of a representation $U_0$ of $\frak g$, then $U_0$ is said
to be an embedded representation.
\medskip

Subrepresentations and linear combinations of equivalent irreps are
trivial examples of embedded representations. Non-trivial
examples are found for semi-direct sum Lie algebras of the
rotor model kind (semi-direct sums with Abelian ideals). Other Lie
algebras contract to this kind of algebra in large quantum number limits
and, consequently, have very good approximations to embedded
representations. The su(3) and symplectic model algebras are examples of
the latter. This is  important for the microscopic theory of collective
motion because, although spin-orbit and other residual interactions 
break the dynamical symmetries of the su(3) and symplectic models, they
mix  representations in a highly coherent way that preserves the algebraic
structures of these models as quasi-dynamical symmetries.
This was predicted to happen as an algebraic expression of an
adiabatic separation of rotational and intrinsic degrees of
freedom\cite{Carv} according to the Born-Oppenheimer approximation. Thus,
it is exciting to discover how extraordinarily good quasi-dynamical
symmetry is in practical situations. 

\section{The rigid rotor algebra as a quasi-dynamical symmetry of the
soft-rotor model}

Without vibrational degrees of freedom, the soft-rotor model is
not an algebraic model. It nevertheless has a  quasi-dynamical
symmetry given by the dynamical symmetry of the (less
realistic) rigid-rotor model.

A spectrum generating algebra for a rigid-rotor model\cite{Ui} is spanned
by three angular momentum operators and five quadrupole moments.
The angular momenta span an so(3) subalgebra;
the quadrupole moments commute among themselves as elements of an Abelian
subalgebra and transform under rotations as components of a rank two
spherical tensor. This algebra, known as rot(3), has irreducible
unitary representations characterized by rigid intrinsic quadrupole shape
parameters $\beta$ and $\gamma$, related to the rotational
invariants by
\bea  [Q\otimes Q]_0 \propto \beta^2 \,,\quad 
[Q\otimes Q\otimes Q]_0 \propto \beta^3\cos 3\gamma \,.
\eea
Rigid-rotor irreps have basis wave functions expressible in the
language of coherent state theory in the form
\be \Psi^{(\beta,\gamma)}_{KLM}(\Omega)
=\langle\beta,\gamma|R(\Omega)|KLM\rangle \,.
\ee

In the physical world, there is no such thing as a truly rigid
rotor. Real rotor wave functions, have intrinsic wave
functions that are linear superpositions of rigid-rotor intrinsic wave
functions with vibrational fluctuations;
\be \Phi_{KLM}(\Omega) = \int
\psi(\beta,\gamma)\,\langle\beta,\gamma|R(\Omega)|KLM\rangle\,
dv(\beta,\gamma) \,.\ee
Due to  Coriolis and centrifugal forces, an intrinsic wave function 
$\psi(\beta,\gamma)$ will generally change with increasing angular
momentum.  However, if the rotational dynamics is adiabatic relative to
the intrinsic vibrational dynamics, then 
$\psi(\beta,\gamma)$ will be independent of $L$ as assumed in the standard
(soft) nuclear rotor model; the rigid-rotor algebra is then  an
exact quasi-dynamical symmetry for the soft rotor. This is clear from the
fact that the matrix elements between states of a soft-rotor model band
are given by
\bea &\lefteqn{\langle \Phi_{K'L'M'} |Q_\nu |\Phi_{KLM}\rangle =
\langle \beta\cos\gamma\rangle 
\int \mathcal{D}^{L'}_{K'M'} (\Omega) \mathcal{D}^2_{0\nu}(\Omega) \mathcal{D}^L_{KM}(\Omega) \, d\Omega }&\cr
&\quad\quad&+\;{1\over \sqrt{2}}\langle \beta\sin\gamma\rangle 
\int \mathcal{D}^{L'}_{K'M'} (\Omega) 
\big[\mathcal{D}^2_{2\nu}(\Omega)+\mathcal{D}^2_{-2,\nu}(\Omega)\big] \mathcal{D}^L_{KM}(\Omega) \, d\Omega \,, \quad
\eea
which is precisely the expression of the rigid-rotor model albeit with
the rigidly-defined values of $\beta\cos\gamma$ and $\beta\sin\gamma$
replaced by their average values.

Note that there is no way to distinguish the states
of a soft-rotor band from those of a rigid-rotor band without considering
states of other bands.
This is because an embedded irrep is mathematically a genuine
representation of the rot(3) algebra; it is simply realized in a way that
may seem contrived from a mathematical perspective but which is
 natural and very physical for a nuclear physicist. Moreover, it is useful
to extract the essence of this   simple  structure because of its
less-than-obvious implications for other dynamical symmetries that have 
rotor and vibrator contractions. 

\section{Effects of the spin-orbit interaction in the SU(3) model}

In molecular physics, one can find near-rigid-rotor spectra of orbital
angular momentum states  weakly coupled by a spin-orbit
interaction to the spins of the atomic electrons.  In nuclear
physics  the spin-orbit interaction is much stronger.  However, far from
destroying the rotational structure of odd nuclei, the spin is usually
strongly coupled to the rotor and participates actively in the formation
of strongly-coupled rotational bands.
Indeed, in the Nilsson model, one includes the spin-degrees of freedom
explicitly in constructing unified model intrinsic states.

\begin{figure}[ht]
\psfig{file=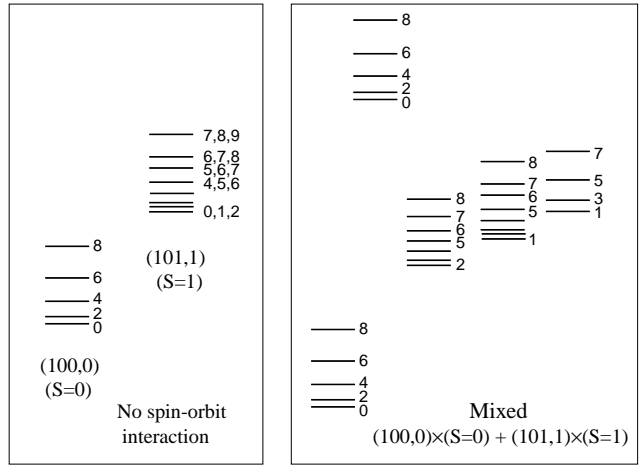,width=3.3in}
%
\caption{The figure shows two SU(3) irreps: a (100,0) irrep with
spin $S=0$ and a (101,1) irrep coupled to states of spin $S=1$. The
right figure shows the result of mixing these two irreps with a strong
spin-orbit interaction
 In spite of the ground and a beta-like
vibrational band being very strongly mixed (essentially 50-50) the
resulting bands would be indistinguishable by experiment from
pure su(3) bands. (The first calculations of this type were carried out by
Rochford\protect\cite{RR88} for lower-dimensional irreps.)
\label{fig:spinorbit}}  
\end{figure}

It is important to recognize that it is
not the spin-orbit interaction that works against strong coupling; it
is the Coriolis force. 
In other words, both a strong rotationally-invariant interaction between
the spin and spatial degrees of freedom and adiabatic rotational motion
(meaning weak centrifugal and Coriolis forces) are important for
strong coupling.
Thus, it was anticipated\cite{Carthesis}
that a spin-orbit interaction might well modify the predictions of a
simple su(3) model and even mix its irreps strongly. But, the underlying
su(3) structure should nevertheless remain discernable and even be
indistinguishable from strongly-coupling rotor model  predictions in the
limit of large-dimensional representations.
In other words, the mixing of su(3) irreps should be highly
coherent as expected for an embedded representation and give 
low-angular momentum states of the form
\be  \Phi_{LM} = \sum_{\lambda\mu} C_{\lambda\mu K}
\Psi_{\lambda\mu KLM}
\ee
with $C_{\lambda\mu K}$ coefficients essentially independent of
$L$. Calculations\cite{RR88}, cf.\ 
Fig.\ \ref{fig:spinorbit}, confirm this to a high degree of accuracy. 
(Note that, because the SU(3) model does
not include Coriolis interactions, the quasi-dynamical symmetry for this
sitation becomes exact in the
$\lambda+\mu\to\infty$ rotor limit.)

\section{SU(3) quasi-dynamical  symmetry and major shell
mixing} 

We know,  from Nilsson model calculations, that major shell
mixing is essential for a reasonable microscopic description of
rotational states.
We also know that, while the symplectic model does not adequately 
account for the spin-orbit and short-range interactions, it contains the
rigid-rotor and quadrupole vibrational algebras as subalgebras and,
consequently, does well as regards the long-range rotational
correlations. Thus, on the basis of many preliminary investigations, we
are confidant that the symplectic algebra, sp$(3,\Bbb{R})$, should be an
excellent quasi-dynamical symmetry for a realistic microscopic theory of
nuclear rotational states. At this time, I show results which
demonstrate that, within a quite large symplectic model irrep with a
Davidson interaction, both the su(3) and rigid-rotor algebras  are
also extraordinarily good quasi-dynamical symmetries. 

\begin{figure}[ht]
\psfig{file=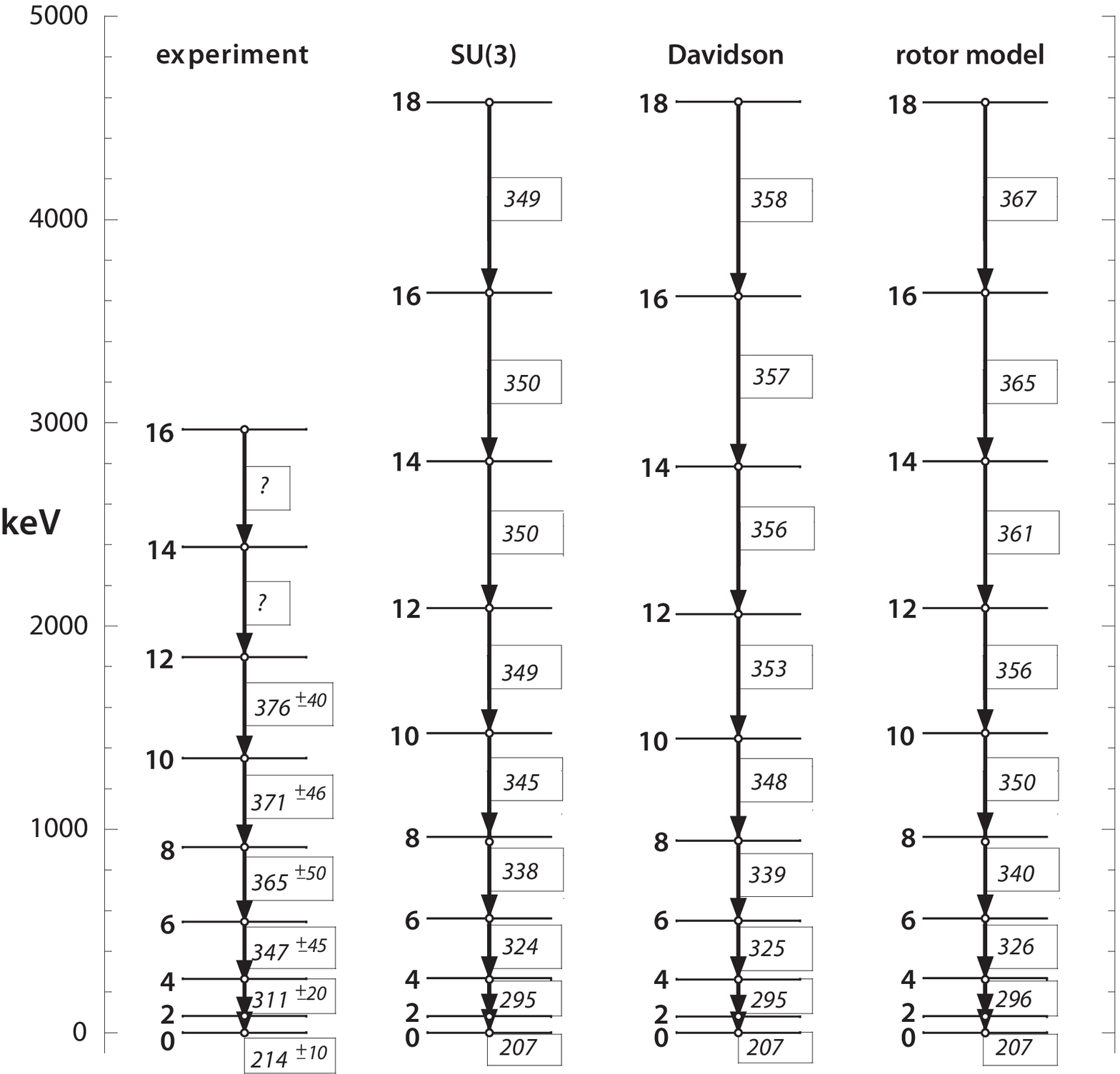,width=3.5in}
%
\caption{Fits to the ground state band of $^{166}$Er with the SU(3),
symplectic, and rigid-rotor models\protect\cite{BR2000}.
\label{fig:SU3Sp3}} 
\end{figure}

Fig.\ \ref{fig:SU3Sp3} shows the spectrum of $^{166}$Er fitted with
three models\cite{BR2000}: the su(3), symplectic, and rigid-rotor 
models, The fitted results are  barely distinguishable; they are
equally successful at fitting the lower levels and E2 transitions and
equally unsuccessful at taking account of centrifugal stretching effects.
In the symplectic model case, this is due to the Davidson potential.

\begin{figure}[ht]
\psfig{file=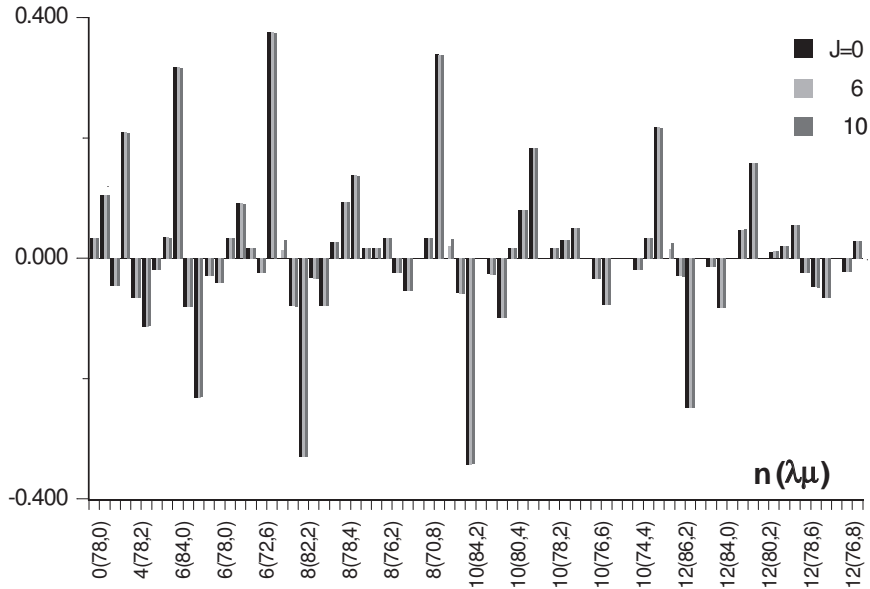,width=4.5in}
\caption{Expansion coefficients of symplectic-Davidson model wave
functions in a multi-shell SU(3) basis covering 12 major harmonic
oscillator shells\protect\cite{BR2000}.
\label{fig:Erwfns}}  
\end{figure}

Fig.\ \ref{fig:Erwfns} shows what the symplectic model wave functions
look like in an su(3) basis.  They exhibit an extraordinary degree of
coherence; i.e., the cofficients are independent of angular  momenta for
a large range of values and indicate the goodness of su(3) as a
quasi-dynamical symmetry.

\nopagebreak\section{SU(3) quasi dynamical symmetry for a model with
pairing interactions}

Finally, we investigated what happens in a 
model that includes both pairing and $Q\cdot Q$ interactions with a
Hamiltonian of the form
\be H(\alpha) = H_0 + (1-\alpha)  V_{\rm su(2)} + \alpha V_{\rm
su(3)}\,, \label{eq:P+Q}
\ee
where  $ V_{\rm su(2)}= -G\hat S_+\hat S_-$ is an su(2) quasi-spin pairing
interaction and $V_{\rm su(3)}=-\chi Q\cdot Q$
 is an su(3) interaction. When $\alpha$ is zero or one, $H$ is easily
diagonalized because of its respective su(2) and su(3) dynamical
symmetries. However, for intermediate values of $\alpha$, diagonalization
of $H$ is a notoriously difficult problem because of the incompatible
nature of su(2) and su(3); they are incompatible in the sense\cite{Incom}
that, within a given harmonic oscillator shell model space, the only space
that is invariant under both su(2) and su(3) is essentially the whole
$S=T=0$ subspace. 

We therefore considered a model having a unitary symplectic dynamical
symmetry, usp(6) (the smallest Lie algebra that contains
both quasispin  su(2) and su(3)  as subalgebras) and
generated large-dimensional usp(6) irreps by
artificially considering particles of large pseudo spin\cite{RBW,BRW}.

\begin{figure}[th]
\psfig{file=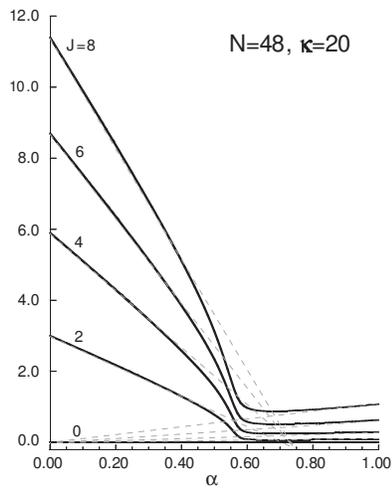,width=2.6in}
\caption{Energy levels of the Hamiltonian (\ref{eq:P+Q})
 as a function of $\alpha$ (taken from ref.\protect\cite{BRW}).
\label{fig:P+Q}}
\end{figure}

\begin{figure}[pt]
\psfig{file=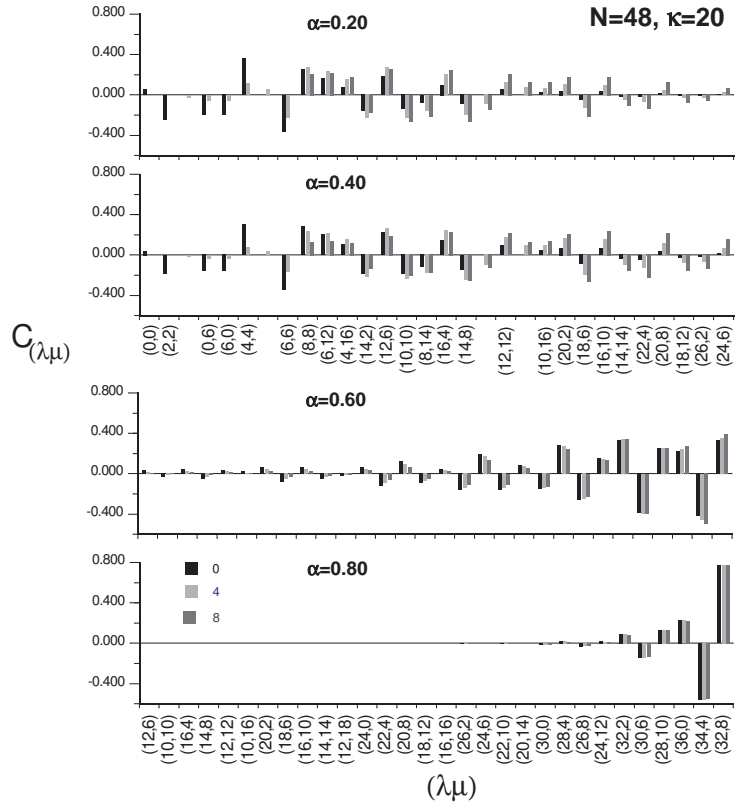,width=4.3in}
\caption{Eigenfunctions of the Hamiltonian (\ref{eq:P+Q}) for four values
of $\alpha$ shown as histograms in an SU(3) basis (taken from
ref.\protect\cite{BRW}).
\label{fig:P+Qwfns}} 
\end{figure}

The  lowest energy states of  $J=0, \dots, 8$ are shown in
Fig.\ \ref{fig:P+Q}.
The results exhibit a phase transition at a critical value of
$\alpha\approx 0.6$ that becomes increasingly sharp as the number of
particles is increased. However, the system does not  flip from
an su(2) to an su(3) dynamical symmetry at the critical point. In
fact, it undergoes a second order phase transition in which the su(3)
symmetry above the critical point is a quasi-dynamical symmetry. This
is seen by looking at the extraordinary coherence of the wave functions
shown for four values of
$\alpha$  in Fig.~\ref{fig:P+Qwfns}. When $\alpha =1$ (not shown) the
wave functions, of course, belong to a single su(3) irrep but, for
smaller values of
$\alpha > 0.6$, they straddle large numbers of su(3) irreps with expansion
coefficients that are essentially independent of angular momentum, as
characteristic of a quasi-dynamical symmetry. 

\nopagebreak

\section{Concluding remarks}

What destroys rotational bands is not the residual interactions.  It is
the Coriolis and centrifugal forces.  Thus,  we can expect
quasi-dynamical symmetry to be a characteristic  of any realistic
description of rotational states.

I conjecture that  quasi-dynamical
symmetry  will prove essential for a realistic
microscopic theory of the rotational states observed in nuclei and
 other many-body systems.
My belief that this will be the case is a response to the fundamental
question:  why do physical many-body systems exhibit rotational bands?
In spite of huge efforts to separate the variables of a many-body system
into subsets of intrinsic and collective variables, the fact 
remains that the separation of collective dynamics is fundamentally due
to the adiabaticity of collective motions (as understood long ago by the
architects of the collective models). Thus, after years of grappling with
the complexity of realizing collective states in microscopic terms, the
conclusion emerges that unless we give the adiabatic principle a central
place in the theory, there is no way we will ever succeed. 
The 
remaining question is: just how do we do this?


\begin{thebibliography}{10}
\bibitem{Carv}  J. Carvalho, R. Le Blanc, M. Vassanji, D.J. Rowe and J.
McGrory, 1986, {\it Nucl.\ Phys.} A{\bf 452}, 240 (1986).

\bibitem{RRR} D.J. Rowe, P. Rochford and J. Repka, {\it J. Math.\ Phys.}
{\bf 29}, 572 (1988).

\bibitem{RR88}  P. Rochford and D.J. Rowe, {\it Phys.\ Lett.} B{\bf 210},
5 (1988).

\bibitem{Ui} H. Ui, {\it Prog.\ Theor.\ Phys.}, {\bf 44}, 153 (1970).

\bibitem{Carthesis} J. Carvalho {\it Ph.D. thesis} (Univ.\ of Toronto,
1984).

\bibitem{LeBCR} R. Le Blanc, J. Carvalho, and D.J. Rowe, {\it Phys.\
Lett.} B{\bf 140}, 155 (1984).

\bibitem{BR2000} C. Bahri and D.J. Rowe, {\it Nucl.\ Phys.} A{\bf 662},
125 (2000).

\bibitem{Incom} D.J. Rowe, ``Compatible and incompatible symmetries in the
theory of nuclear collective motion'', in {\it  New Perspectives in
Nuclear Structure\/} (ed.\ Aldo Covello, World Scientific) pp 169-183.

\bibitem{RBW} D.J. Rowe, C. Bahri and W. Wijesundera, {\it Phys.\ Rev.\
Lett.}, {\bf 80}, 4394 (1998).

\bibitem{BRW} C. Bahri, D.J. Rowe, and W. Wijesundera, {\it Phys.\ Rev.}
C{\bf 58}, 1539 (1998).

\end{thebibliography}
\end{document}